\documentclass{article}

\usepackage[english]{babel}

\usepackage[letterpaper,top=2cm,bottom=2cm,left=3cm,right=3cm,marginparwidth=1.75cm]{geometry}

\usepackage{amsmath}
\usepackage{graphicx}
\usepackage[colorlinks=true, allcolors=blue]{hyperref}

\title{Reviewing the GR Method for Estimating Black Hole
Parameters of Megamaser Systems}
\author{Adriana González-Juárez$^{*}$ and Alfredo Herrera-Aguilar\vspace{16pt}\\
Instituto de F\'isica, Benem\'erita Universidad Aut\'onoma de Puebla,\vspace{3pt}\\
Apdo. Postal J-48, CP 72570, Puebla, M\'exico}

\begin{document}
\maketitle

\begin{abstract}
We review a General Relativistic (GR) method to determine the black hole (BH) parameters: mass-to-distance ratio, position and recessional velocity of active galactic nuclei (AGNs) of Seyfert type, which have an accretion disk with water masers circulating around the BH. This GR method makes use of astrophysical observations: the redshifted and the blueshifted photons emitted from the aforementioned masers and their orbital position on the sky. In order to perform the estimations we implement a Bayesian statistical method to fit the above mentioned observational data. One of the main results of this work consists in analytically expressing the gravitational redshift, allowing us to quantify its magnitude for the photons emitted by the closest masers to the black holes. We present this quantity for several BHs hosted at the core of AGNs.
\end{abstract}

\textcolor{white}{falsespace}\\
$^*$adrianag@ifuap.buap.mx\\

\textcolor{white}{falsespace}\\
\textbf{Keywords}: black hole parameters - general relativity - megamaser systems.

\section{Introduction}

An interesting process in galactic dynamics regards the mechanisms that lead to the creation of supermassive  black holes (BHs) in the center of most galaxies (see, e.g., Kormendy \& Richstone \cite{Kormendy1995}).
The current understanding that supermassive BHs are found at the core of many galaxies arose with the discovery of quasars and AGNs, where huge rates of luminous radiation are produced within small compact volumes (see Celotti et al. \cite{Review}). Besides, the presence of a supermassive BH at the core of these AGNs plausibly explains the spectral information coming from gas particles that revolve around it and attain velocities of thousands of kilometers per second. 
A relevant issue regarding the characterization of AGNs consists in determining the mass, the spin, the distance as well as  other parameters of the BHs hosted at their cores. In particular, determining the BH masses in AGNs is essential for understanding the fundamental properties of their central engines, as well as for gaining insights into their growth and coevolution with their host galaxy (see, e.g., Kormendy \& Ho \cite{Kormendy2013}, Greene et al. \cite{Green2016}).

Water megamaser systems consist of H$_2$O vapor clouds that emit intense stimulated microwave radiation at 22 GHz and offer a powerful tool for investigating the very core of AGNs and their surrounding environments. These astrophysical systems are very luminous water masers, typically $\sim\!\!10^6$ times more luminous than Galactic maser sources, hosted on accretion disks at circumnuclear regions of several AGNs. Their extremely high surface brightness enables a detailed mapping at sub-milliarcsecond resolution using Very Long Baseline Interferometry (VLBI), allowing for a direct way to study AGN structures and dynamics at sub-parsec scales.

On the other hand, the maser disks themselves provide valuable insights into the dynamics around supermassive BHs. In particular, the maser disk size correlates with the central BH mass: The mean radius of maser disks increases with BH mass (see Wardle \& Yuset-Zadech \cite{Wardle}, Gao et al. \cite{MCPIX}, Kuo et al. \cite{DiskSize}). Besides, according to a recent model developed by Kuo et al. \cite{DiskSize}, maser disks also exhibit interesting behavior related to their size, which is influenced by physical processes and the interplay of various factors. Namely, the outer radius of the disk is mainly defined by the maximum X-ray heating rate coming from the central engine or by the minimum gas density required for efficient maser emission, depending on a combination of the Eddington ratio, the BH mass, and the disk mass. Meanwhile, the disk inner radius for maser action is determined by the dust sublimation radius.

Within this context, the Megamaser Cosmology Project\footnote{The MCP is a key project of the National Radio Astronomy Observatory (NRAO), in collaboration with the Cosmic Microwave Background project from Wilkinson Microwave Anisotropy Probe and Planck missions,
that seeks to determine the Hubble constant value by making use of megamaser observations of galaxies moving within the Hubble flow, Reid et al. \cite{MCPI}. \url{https://safe.nrao.edu/wiki/bin/view/Main/MegamaserCosmologyProject}
} 
(MCP) has studied over 20
megamaser disks in AGNs (see, for instance, Reid et al. \cite{MCPI}, Kuo et al. \cite{MCPIII}, Kuo et al. \cite{MCPV}, Gao et al. \cite{MCPVIII}, Gao et al. \cite{MCPIX}, Zhao et al. \cite{MCPX}, Pesce et al. \cite{MCPXI}, Kuo et al. \cite{MCPXII}, and Pesce et al. \cite{MCPXIII}) making use of VLBI techniques with the Very Large Baseline Array (VLBA), the Radio Telescope Effelsberg (ET), the Robert C. Byrd Green Bank Telescope (GBT) and the Karl G. Jansky Very Large Array (VLA). 
This project has managed to determine the Hubble constant value $H_0$ with $4\%$ uncertainty (see Pesce et al. \cite{MCPXIII}), and to estimate the mass of the BH hosted at the center of those galaxies to percent-level accuracy (see Kuo et al. \cite{MCPIII}, Gao et al. \cite{MCPIX}) as well as the distance to them, disregarding the use of distance ladders, gravitational lenses or standard candles (see Reid et al. \cite{MCPI}, Gao et al. \cite{MCPVIII}, Pesce et al. \cite{MCPXI}).

Within this framework, in order to determine the parameters of the  supermassive BHs located at the core of AGNs, a full general relativistic method was developed in Herrera-Aguilar \& Nucamendi \cite{Herrera15} and Banerjee et al. \cite{Banerjee22} for spinning BH configurations, allowing for the potential detection of relativistic effects within such astrophysical systems.

A static version of this method was applied to estimate the mass-to-distance ratio\footnote{This ratio refers to the BH mass $M$ divided by its distance to the observer $D$.} 
of the central BHs living in several AGNs: The case of NGC 4258 was addressed in Nucamendi et al. \cite{Nucamendi21}, whereas TXS-2226-184 was studied in Villalobos-Ram\'irez et al. \cite{Artemisa22}; fourteen more galaxies previously considered by the MCP were approached in Villaraos et al. \cite{Villaraos} and Gonz\'alez-Ju\'arez et al. \cite{Adriana2024}. In these works, the authors also quantified general and special relativistic effects, namely, the gravitational redshift (generated by the curvature of spacetime in the BH vicinity) experienced by the closest maser to the BH, and the peculiar velocity of the host galaxy with respect to the Earth, accounted for by a special relativistic boost.

This work is organized as follows: In section \ref{sec:GR} we present the BH rotation curve general relativistic model, whereas in Sec. \ref{sec:bayes} we briefly review the Bayesian statistical method that we apply to it and present the results of the fit. Finally, we conclude 
in Sec. \ref{sec:conclusions2}. 

\section{Frequency shift in general relativity }
\label{sec:GR}

In this Section we review the general relativistic model that we employ to describe BH rotation curves. This model is based on a metric approach developed in Herrera-Aguilar \& Nucamendi \cite{Herrera15} and Banerjee et al. \cite{Banerjee22} that makes use of observational quantities, namely, the redshift and blueshift of photons emitted by massive particles orbiting the BH and their orbital parameters. 

This method was also implemented in Nucamendi et al. \cite{Nucamendi21} using a Schwarzschild metric in order to fit the parameters of the BH hosted at the core of the NGC 4258 AGN. In this work, water megamaser clouds that lie on the accretion disk and circularly revolve around the central BH of the AGN were considered as test particles that are stimulated by the BH, making them to emit photons in a very coherent way. Both the maser clouds geodesically orbiting the BH and the photons they emit feel the gravitational field of the BH and keep memory of its properties. Thus, when we measure on Earth the shift in the photon's frequency at certain orbital positions of the masers, we obtain information about the BH parameters as well. 

We further derive the general relativistic formulas for the total frequency shift experienced 
by photons emitted by massive bodies orbiting a  receding or approaching Schwarzschild BH with respect to a distant observer with constant peculiar velocity $v_p$.
We also expand these expressions in terms of the $m/r_e$ and $v_p/c$ ratios in order to see what is the magnitude of the general and special relativistic corrections to the corresponding Newtonian formula. 
We finally compose the frequency shifts generated by the special relativistic boost and the 
expansion of the Universe in order to account for the recessional redshift of BHs hosted at the core of galaxies within the Hubble flow.

\subsection{Geodesic motion of massive particles around the Schwarzschild BH}

We start by recalling the expression
for the Schwarzschild metric
\begin{equation}
ds^{2}=\dfrac{dr^{2}}{f}+r^{2}(d\theta ^{2}+\sin {\theta }^{2}d\varphi
^{2})-fdt^{2},\quad \quad f=1-\frac{2m}{r},  \label{schw}
\end{equation}%
where $m$\ is the total mass of the BH in geometrized units ($G=1=c$). The motion of massive and massless particles takes place in the gravitational field of this line element.

The per mass unit conserved energy $E$ and axial angular momentum $L$ of a massive particle in geodesic motion in the Schwarzschild background due to the existence of the temporal $ \xi^\mu=\delta_t^\mu$ and rotational $\psi^\mu=\delta_{\varphi}^\mu$ Killing vector fields read
\begin{eqnarray}
  E = - g_{\mu\nu} \xi^{\mu} U^{\nu} = - g_{tt}U^{t} , \qquad
  L =  g_{\mu\nu} \psi^{\mu} U^{\nu} = g_{\varphi\varphi} U^{\varphi}, \qquad 
\end{eqnarray}
where $ g_{\mu\nu}$ is the metric tensor and $U^{\mu}$ is the particle four-velocity.

These relations allow us to express the four-velocity components $U^{t}$ and $U^{\varphi}$ as follows
\begin{eqnarray}
  U^{t} = -\frac{ E }{g_{tt}}, \qquad \qquad
  U^{\varphi} = \frac{L}{g_{\varphi\varphi}}.
\end{eqnarray}
The momentum conservation in GR restricts the four-velocity to $U^2=-1$, giving
\begin{eqnarray}
    g_{rr} (U^{r})^2 + g_{\theta\theta} (U^{\theta})^2 + \frac{E^2}{g_{tt}} + \frac{L^2}{ g_{\varphi\varphi}} + 1 = 0,
\end{eqnarray}
Since the Schwarzschild metric possesses spherical symmetry, we can always restrict particle motion to the equatorial plane $\theta = \pi/2$ ($U^{\theta}=0$), obtaining the following non-relativistic energy conservation equation
\begin{eqnarray}
    \frac{1}{2} (U^{r})^2 +  \frac{1}{2} \left( 1- \frac{2m}{r} \right) \left(\frac{L^2}{ r^2} + 1 \right) = \frac{E^2}{2},
\end{eqnarray}
which defines an effective potential of the form
\begin{eqnarray}
   V_{eff} = \frac{1}{2} \left( 1- \frac{2m}{r} \right) \left(\frac{L^2}{ r^2} + 1 \right).
   \label{Veff}
\end{eqnarray}
By further considering circular motion ($U^r=0$), we require to have a minimum in the effective potential, obtaining the following conditions
\begin{eqnarray}
   V_{eff} = \frac{E^2}{2}, \qquad \qquad \partial_r V_{eff} = 0.
   \label{dVeff}
\end{eqnarray}
From these relations we obtain $E$ and $L$ as follows:
\begin{eqnarray}
E = \frac{1- \frac{2m}{r} }{\sqrt{1 - \frac{3m}{r}} }, \qquad
L = \pm \,r \sqrt{\frac{\frac{m}{r}}{1 - \frac{3m}{r}}}.
\end{eqnarray}
where the $\pm$ signs correspond to clockwise and counterclockwise motion of the test particles with respect to a distant observer. These relations are very interesting since they point out that both energy and angular momentum of massive particles in circular geodesic motion are divided by the factor $\sqrt{1 -3m/r}$.

Therefore, the non-trivial components of the four-velocity read
\begin{eqnarray}
U^t = \frac{1}{\sqrt{1 - \frac{3m}{r}} }, \qquad
U^{\varphi} = \pm \,\frac{1}{r} \sqrt{\frac{\frac{m}{r}}{1 - \frac{3m}{r}}}.
\end{eqnarray}

It is also crucial to note that the energy of a massive particle in circular geodesic motion differs from that of a {\it static} particle in the background of a Schwarzschild BH (see Schutz \cite{Schutz09}). Moreover, a static particle cannot be in geodesic motion in the sense that it would necessarily move towards the BH. Therefore, in order to model general relativistic effects such as the gravitational frequency shift or the time dilation of particle's motion on accretion discs in a correct way, one has to consider massive particles circularly orbiting the central BH.

\subsection{Geodesic motion of photons in the Schwarzschild metric}

We now turn to describe the motion of photons emitted by massive particles in the Schwarzschild BH metric. The propagation of photons is parameterized by their four-momentum 
$k^{\mu}$. By following the same line of reasoning used for massive particles, we shall have two conserved quantities, energy and axial angular momentum, along the photons path due to the existence of the temporal and rotational Killing vector fields:
\begin{eqnarray}
E_{\gamma} = - g_{tt}k^{t}, \qquad
L _{\gamma} = g_{\varphi\varphi} k^{\varphi}.
\end{eqnarray}
These relations allow us to express the $k^t$ and $k^{\varphi}$ components as
\begin{eqnarray}
k^{t} = - \frac{E_{\gamma}}{g_{tt}} = \frac{E_{\gamma}}{ 1- \frac{2m}{r} }, \qquad
k^{\varphi} = \frac{L _{\gamma}}{g_{\varphi\varphi}} = \frac{L _{\gamma}}{r^2}.
\end{eqnarray}
Moreover, the photon paths are restricted to the condition $k^2=0$, which on the equatorial plane yields an expression for $k^r$:
\begin{equation}
k^r = \sqrt{E_{\gamma}^2 - \left(1- \frac{2m}{r} \right)\frac{L _{\gamma}^2}{r^2} },
\end{equation}
completely determining all the nontrivial components of the photon's four-momentum since the $k^{\theta} = 0$ due to the motion on the equatorial plane.

\subsection{The frequency shift in Schwarzschild background}

We further define the photon frequency as a general relativistic invariant $\omega=-k_{\mu}U^{\mu}$. This quantity can be measured at the points of emission and detection, allowing us to arrive at the following expression for the Schwarzschild frequency shift of these photons in the equatorial plane
\begin{eqnarray}
1 + z_{Schw_{1,2}}= \frac{\omega_e}{\omega_d} = \frac{\left(E_{\gamma} U^t - L_{\gamma} U^{\varphi} - g_{rr} k^rU^r \right)_e }{\left(E_{\gamma} U^t - L_{\gamma} U^{\varphi} - g_{rr} k^rU^r \right)_d} = \frac{\left( U^t - b_{\mp} U^{\varphi} \right)_e }{\left( U^t - b_{\mp} U^{\varphi} \right)_d},
\label{zSchw1,2}
\end{eqnarray}
where the indices $_{1,2}$ refer to redshift and blueshift, respectively, while the subindices $_{e}$ and $_{d}$ denote an emission and detection points. Here we have considered that the emitter and the detector are in geodesic circular motion around the BH, and

we have defined the light bending parameter (also called apparent impact parameter) at the points at which the velocity gain paths of the test particles are the longest (around the disk midline) in order to maximize the possibilities of frequency shift detection:
\begin{equation}
b_{\pm} \equiv \frac{L_{\gamma}}{E_{\gamma}} =  \pm \sqrt{-\frac{g_{\varphi\varphi}}{g_{tt}}} = \pm \frac{r}{\sqrt{1 - \frac{2m}{r}}},
\end{equation}
where now the $\pm$ signs correspond to the massive particle's motion either side of the observer's line of sight.
This quantity measures in fact the deflection of light generated by the gravitational field of the BH, it is conserved from the moment of emission till detection ($b_e=b_d$), and depends purely on the metric.

By considering a static observer located far away from the BH, i.e. when $r_d \longrightarrow \infty$, its four-velocity simplifies to
$
\left.U^{\mu}\right|_d = (1,0,0,0),
$
rendering the following total frequency shift 
\begin{equation}
1 + z_{Schw_{1,2}} = \frac{1}{\sqrt{1 - \frac{3m}{r_e}}}\left(1 \pm \sqrt{\frac{\frac{m}{r_e}}{1 - \frac{2m}{r_e}}} \right),
\label{zSchw}
\end{equation}
where the $\pm$ signs correspond to the $_{1,2}$ indices and refer to the redshift and the blueshift, respectively, and $r_{e}$\ is the orbital radius of the emitter (the megamaser features). Here it is suitable to use the approximation $ \Theta \approx r_e / D$, where $\Theta$ is the angular distance between a given maser and the BH in our model and estimations. 

From this expression, it is quite natural to define the gravitational redshift in terms of the temporal component of the contraction that defines the Schwarzschild frequency shift (\ref{zSchw1,2}), a quantity that encodes the curvature of spacetime generated by the BH mass and has no Newtonian analogue:
\begin{equation}
1 + z_g = \frac{1}{\sqrt{1 - \frac{3m}{r_e}}},
\label{zg}
\end{equation}
whereas the kinematic redshift and blueshift either side of the BH are determined by the second term:
\begin{equation}
z_{kin_{\pm}} = \pm \sqrt{\frac{\frac{m}{r_e}}{\left( 1 - \frac{3m}{r_e} \right) \left(1 - \frac{2m}{r_e}\right)}}
\label{zkin}
\end{equation}
that also modifies the Newtonian kinematic frequency shift. 

We remark that in this work we only use the redshift and blueshift of masers located at the points where their velocity gain paths are the longest, i.e., in the vicinity of the midline of the disk. The use of systemic maser features still remains pending.

In the weak field limit, by expanding $z_{Schw_{1,2}} $ with respect to $m/r_e$ we see that the leading term is given by the redshift and blueshift of photons corresponding to rotational motion in the Newtonian picture, whereas the next-to-leading term renders a general relativistic correction due to the gravitational redshift
\begin{equation}
z_{Schw_{1,2}} \approx 
z_{kin_{\pm}} + z_g + \cdots =
\pm \sqrt{\frac{m}{r_e}} + \frac{3}{2}\frac{m}{r_e} + \cdots ;
\label{expzSchw}
\end{equation}
here the first term constitutes the first order approximation of the kinematic frequency shift (\ref{zkin}), while the second item represents the first order approximation of the gravitational redshift (\ref{zg}).

\subsection{A receding or approaching BH and the total redshift}

Now we consider that the Schwarzschild BH is locally receding from or approaching to the distant observer as a whole entity with a constant peculiar velocity $v_p$. This local motion can be described by a special relativistic boost since it is not generated by the Universe's expansion and produces, in turn, a special relativistic redshift or blueshift that we call $z_{boost}$ with the following definition (see Rindler \cite{RindlerSR1989}): 
\begin{equation}
1 + z_{boost} = \gamma (1 + \beta \cos\kappa), \quad \gamma=1/\sqrt{1-\beta^2},  \quad \beta=v_p/c, 
\label{z_boost}  
\end{equation}
where $\kappa$ is the angle between the direction of the peculiar velocity and the LOS (we shall consider $\kappa=0$ henceforth for simplicity), $v_p \equiv z_p c$, and we have called $z_p$ the redshift or blueshift expressed in terms of the peculiar velocity $v_p$ of the galaxy hosting the BH. This quantity is independent of the BH mass and has a different nature compared to the Schwarzschild redshift since it is generated by a special relativistic effect, i. e. by a change of reference frame moving with velocity $v_p$ with respect to the reference frame of a distant observer in flat spacetime.

By further composing the Schwarzschild and the special relativistic frequency shifts according to Davis \& Scrimgeour \cite{Davis14} we obtain the local redshift: 
\begin{equation}
1+z_{loc_{1,2}} = (1+z_{Schw_{1,2}})(1 + z_{boost}).
\label{composez}
\end{equation}
The local redshift (or bueshift), $z_{loc_{1,2}}$, is the quantity that one measures from a photon's source revolving a receding or approaching BH along with its position on the sky.

By performing a double expansion of the local frequency shift in terms of the $m/r_e$ ratio and the special relativistic boost parameter $v_p/c$, we obtain up to $1.5$ order  (see Nucamendi et al. \cite{Nucamendi21})
\begin{eqnarray}
z_{tot_{1,2}} & \approx & 
z_{kin_{\pm}} + z_p + z_g + z_p \, z_{kin_{\pm}} + z_{kin_{\pm}} \, z_g + \cdots \nonumber
\\
& = &\pm \sqrt{\frac{m}{r_e}} + \frac{v_p}{c} + \frac{3}{2}\frac{m}{r_e} \pm \frac{v_p}{c} \sqrt{\frac{m}{r_e}} \pm \frac{5}{2} \left(\frac{m}{r_e}\right)^{\frac{3}{2}} + \cdots , 
\end{eqnarray}
where additionally to the items obtained in the expansion (\ref{expzSchw}), now the second term corresponds to the receding or approaching peculiar velocity of the BH with respect to a distant observer; the fourth term constitutes a combined effect of kinematic and boost redshifts, called kinematic boosted redshift, and the fifth term denotes a composed effect of the gravitational redshift with the kinematic frequency shift; finally, ellipses stand for higher order contributions that become important when the orbiting objects are very close to the BHs.

\subsection{The cosmological redshift due to Universe's expansion}

In order to have a more realistic modelling of the recessional redshift, ${z_{rec}}$, of BHs hosted at galactic cores that are within the Hubble flow we also need to take into account the cosmological redshift, $z_{cosm}$, i.e. the
stretching of photons' wavelength produced by the expansion of the Universe. 

Thus, the recessional redshift accounting for both local and cosmological motion of galaxies within the Hubble flow reads
\begin{equation}
1+z_{rec}=(1+z_{boost})(1+z_{cosm}),
\label{z_rec}
\end{equation}
where the cosmological redshift depends on the metric chosen to describe the expansion of the Universe.

However, the Schwarzschild metric we use in our model is static and does not provide information about the Universe's expansion.
On the other hand, the special relativistic  frequency shift (\ref{z_boost}) does not depend on the metric either, leading to a degeneracy of the cosmological and the peculiar redshifts when performing statistical estimations of BH parameters using astrophysical observational data (since none of them depend on the metric). Therefore, in order to avoid this degeneracy, in this work we will fit the total recessional redshift (\ref{z_rec}) instead of its separate components.

Thus, the expression for the total redshift of photons composing the Schwarzschild frequency shift and the recessional redshift as a first approximation reads 
\begin{equation}
1+z_{tot_{1,2}}=(1+z_{Schw_{1,2}}){(1+z_{rec})}.  
\label{redshi_tot}
\end{equation}
This is the formula that we shall use in order to determine the BH parameters using Bayesian statistical fits of observational data.

\section{Bayesian statistical model}
\label{sec:bayes}

In order to further implement the general relativistic method to fit real megamaser astrophysical data, we need to develop a Bayesian statistical model that takes into account the positions and velocities of redshifted and blueshifted water masers (located close to the midline) along with their corresponding uncertainties.
This statistical method is based on a Monte Carlo method that makes use of Markov chains and consists of a least squares $\chi ^{2}$ fit of the following parameters: the BH mass-to-distance ratio, $M/D$, its recessional velocity, $z_{rec}$, and its position on the sky (either along just the $x-$offset or both the $x-$ and $y-$offsets).
In a real maser map, the detected spots do not lie perfectly on the midline and are spread around it. Therefore, following an original idea of Herrnstein et al. \cite{Herrnstein05}, we introduce into our model a small dispersion in the azimuthal angle $\varphi-\varphi_0$ that encodes the departure of the redshifted and blueshifted masers from a fixed value $\varphi_0$ (we set $\varphi_0=0$ corresponding to the midline). We fix the amplitude in the scattering angle by choosing the smallest value that renders a reduced $\chi ^{2}$ close to unity.
Thus, the Bayesian statistical model is given by
\begin{equation}
\chi ^{2}=
\sum_{k=1}\frac{\left[ {z_{k,obs}}-(1+z_{g}+
{\sin {%
\theta _{0}}}~
\cos\varphi~
z_{kin_{\pm }}){(1+z_{rec})}+1\right] ^{2}}{\sigma
_{z_{tot_{1,2}}}^{2}+
\sin ^{2}{\theta _{0}}~z_{kin_{\pm }}^{2}{%
(1+z_{rec})^{2}}
\sin^2\varphi~\delta\varphi^2}, 
\label{chi}
\end{equation}
where $\sigma_{z_{tot1,2}}=|\delta z_{tot_{1,2}}|$ is the uncertainty of the total redshift, $\theta_0$ stands for the inclination angle, 
$\delta\varphi$ stands for the induced uncertainties of the maser spread and we have assumed small variations $\delta\varphi \ll 1$.

The error of the total redshift $\sigma
_{z_{tot_{1,2}}}^{2}$ has the following form:
\begin{equation}
\delta z_{tot_{1,2}}=(\delta z_{g}+\delta z_{kin_{\pm }})(1+z_{rec}),
\end{equation}%
where
\begin{equation}
\delta z_{g}=\left( 1+z_{g}\right) ^{3}\left( \dfrac{-3m}{%
2r_{e}}\right) \dfrac{\delta r_{e}}{r_{e}}, \qquad \delta r_e\approx D~\delta\Theta, 
\end{equation}%
\begin{equation}
\delta z_{kin_{\pm }}= \sin {\theta _{0}}\cos\varphi\left( z_{kin_{\pm
}}\right) ^{3}\left( \dfrac{6m^{2}-r_{e}^{2}}{2mr_{e}}\right) \dfrac{\delta r_{e}}{r_{e}}, 
\end{equation}%
\begin{equation}
\delta \Theta = \sqrt{\left( \frac{x_{i}-x_{0}}{\Theta}\right) ^{2}\delta
_{x}^{2}+\left( \frac{y_{i}-y_{0}}{\Theta}\right) ^{2}\delta _{y}^{2}},
\end{equation}
\begin{equation}
\Theta = \sqrt{\left( x_{i}-x_{0}\right) ^{2}+\left( y_{i}-y_{0}\right) ^{2}},
\end{equation}
with ($x_i$, $y_i)$ denoting the position of the $i$-th megamaser on the sky, $\{\delta _{x},\delta _{y}\}$ being their corresponding errors, and $(x_{0},y_{0})$ standing for the BH position.

\section{Conclusions}
\label{sec:conclusions2}
In this work we have made a detailed description of the General Relativistic method developed to characterize the main parameters of a compact object orbited by test particles in geodesic motion that emit frequency shifted photons towards a distant observer. We also described how the method is applied to real astrophysical systems using observations, particularly to megamasers in the accretion disks orbiting the AGN’s central black holes. The method is presented as an alternative to the classical Newtonian treatments applied so far.  It is worth mentioning that one of the virtues of the general relativistic method used to estimate the parameters of BHs hosted in the core of AGNs is that it allows for the clear identification of the relativistic effects present in their dynamics. In particular, it enables the gravitational redshift of each highly redshifted or blueshifted maser feature to be quantified. This fact eases the potential detection of such a general relativistic effect in this kind of astrophysical systems. Therefore, as an application of this formalism, we present the gravitational redshift of the closest masers to the BHs located at the center of several galaxies and display the corresponding results in Table \ref{Tabla_zgrav}.

\section*{Acknowledgements}
The authors are grateful to D. Villaraos, U. Nucamendi and M. Momennia for illuminating discussions.
The authors acknowledge CONAHCYT for support under grant No. CF-MG-2558591, VIEP-BUAP, as well as SNII. A.G.-J. was supported by a postdoctoral grant 
under the CONAHCYT program {\it Estancias Posdoctorales por M\'exico 2022}.


\begin{table}
\caption{Sample of studied megamaser systems and the gravitational redshift of the closest maser to the central black holes.}
\label{Tabla_zgrav}
\begin{center}
\footnotesize
\begin{tabular}{cccc}
Source & Distance to & $z_{g}$ ($10^{-6}$) & $M/D$ \\ 
& closest maser & $v_g$\,($km\  s^{-1}$) & $\times 10^5$ \\ 
& (mas) &  & $(M_{\odot}Mpc^{-1}$) \\ \hline\hline
& & & \\
NGC 4258$^1$ & 3.17 & 24.87 & $53.26 \pm 0.02$\\
& & 7.45 & \\ 
\hline\\
NGC 5765b$^2$ & 0.537 & 10.259 & $3.727\pm0.013$ \\
& & 3.075 &  \\ 
\hline\\
NGC 6323$^2$ & 0.215 & 6.310 & $0.916\pm 0.005$\\
& & 1.891 &  \\ 
\hline\\
UGC 3789$^2$ & 0.305 & 11.042 & $2.277\pm0.009$\\
& & 3.310 & \\ 
\hline\\
CGCG 074-064$^2$ & 0.252 & 16.098 & $2.749^{+0.017}_{-0.015}$
\\
& & 4.826 &  \\ 
\hline\\
ESO 558-G009$^2$ & 0.380 & 6.199 & $1.594^{+0.025}_{-0.026}$
\\
& & 1.858 &  \\ 
\hline\\
NGC 2960$^2$ & 0.291 & 8.830 & $1.738^{+0.015}_{-0.016}$\\
& & 2.647 &  \\ 
\hline\\
NGC 6264$^2$ & 0.336 & 9.357 & $2.126\pm0.007$\\
& & 2.805 &  \\ 
\hline\\
J0437+2456$^2$ & 0.140 & 4.476 & $0.425^{+0.010}_{-0.012}$\\
& & 1.341 & \\ 
\hline\\
NGC 4388$^2$ & 1.829 & 3.422 & $4.233^{+0.185}_{-0.268}$\\
& & 1.026 &  \\ 
\hline\\
NGC 2273$^2$ & 0.211 & 22.927 & $3.283^{+0.045}_{-0.051}$\\
& & 6.873 &  \\ 
\hline\\
NGC 1194$^3$ & 2.549 & 7.610 & $13.100^{+0.211}_{-0.209}$\\
& & 2.282 &  \\ 
\hline\\
NGC 5495$^3$ & 0.189 & 9.036 & $1.153^{+0.208}_{-0.174}$\\
& & 2.709 &  \\ 
\hline\\
Mrk 1029$^3$ & 0.460 & 0.465 & $0.144 \pm 0.011$\\
& & 0.139 &  \\ 
\hline\\
NGC 1320$^3$ & 0.809 & 2.740 & $1.497^{+0.071}_{-0.069}$\\ 
& & 0.821 &  \\ 
\hline\hline
\end{tabular}%
\end{center}
The superscript numbers in the first column are related to the following references: 
1 - Nucamendi et al. \cite{Nucamendi21}, 
2 - Villaraos et al. \cite{Villaraos} and
3 - González-Juárez et al. \cite{Adriana2024}. 
\end{table}

\end{document}